\documentstyle[12pt]{article}
\title{
\large
     SUPERSYMMETRY OF A NONSTATIONARY PAULI EQUATION}
      
\author{V. M. Tkachuk \\
Ivan Franko Lviv State University, Chair of Theoretical Physics \\ 
		 12 Drahomanov Str., Lviv UA--290005, Ukraine\\
		 E-mail: tkachuk@ktf.franko.lviv.ua }

\begin{document}
 
\setlength{\jot}{1em}
\setlength{\abovedisplayskip}{1.5em}
\setlength{\belowdisplayskip}{1.5em}

\setcounter{page}{1}

\maketitle
\begin{abstract}
{\small The supersymmetry of the electron in both the nonstationary 
magnetic and electric fields  in a two-dimensional case is studied. 
The supercharges which are the integrals of motion and their algebra
are established. Using the obtained algebra the solutions of nonstationary
Pauli equation are generated.} 

\end{abstract}

PACS number(s): 03.65.-w; 11.30.Pb \\
\pagebreak

\section{Introduction}
Supersymmetric quantum mechanics introduced in the papers \cite{Wit,Nic}
currently attaches much attention (see also surveys \cite{GK,Coop,Jun}).
One of the quantum mechanical problem where supersymmetry (SUSY) 
is the physical symmetry is the motion of the electron in a stationary
two-dimensional magnetic fields and in some three-dimensional
fields
\cite{Grum,Jac,Khare,Gen,Tk}
(see also \cite{GK,Coop,Jun} and references therein).

The problem of introducing supersymmetry in the case of
the electron motion in the nonstationary magnetic field 
was investigated for the first time in \cite{Kost}. 
The case of a time-varying uniform magnetic
field with the fixed direction was examined and 
it was shown that the group-theoretical
analysis can provide a clue to the supersymmetric factorization
of the time-dependent Pauli equation and to the obtaining of eigensolutions.
Note also  that recently the supersymmetry of a one-dimensional 
time-dependent Schr\"{o}dinger equation was established and with the
help of the time-dependent Darboux transformation new exactly
solvable time-dependent potentials were generated \cite{Bag}. 

The aim of the present paper is to establish the supersymmetry
of the electron in both the nonstationary magnetic and electric fields. 
The supercharges are obtained as integrals of motion. 
They are the straightforward generalization
of the well-known supercharges in the stationary case. 

\section{SUSY the of Pauli equation in a two-dimensional case}
In the present paper we consider the electron motion in
nonstationary "two-di\-men\-si\-o\-nal" electromagnetic field. 
The equation of
motion in this case is the nonstationary Pauli equation 

\begin{eqnarray}
&& i{\partial \psi \over \partial t}=H_p \psi,
\end{eqnarray}
where
\begin{eqnarray} 
&& H_p={1 \over 2}(\pi_x^2+\pi_y^2+p_z^2-eB\sigma_z), \\
&& \pi_\alpha=p_\alpha-eA_\alpha, \qquad p_\alpha=-i{\partial \over 
\partial x_\alpha}, \nonumber 
\end{eqnarray}

$$A_x=A_x(x,y,t), A_y=A_y(x,y,t), A_z=0$$ are the components of 
the vector potential,

$$B=B_z(z,y,t)={\partial A_y \over \partial x}-{\partial A_x \over 
\partial 
y}$$ is the magnetic field parallel to the $z$-axis.

It is obvious that  $p_z$ is the integral of motion. Therefore, the
solution of equation (1) can be written in the following form 
\begin{eqnarray}
&& \psi(x,y,z,t)=e^{-ik_z^2t/2+ikz}\psi(x,y,t) \nonumber,
\end{eqnarray}
where $k$ is the value of momentum along $z$-axis. Then $\psi(x,y,t)$ 
satisfies the Pauli equation (1)  with a two-dimensional Hamiltonian
\begin{eqnarray} \label{Pauli_hamiltonian}
&& H={1 \over 2}(\pi_x^2+\pi_y^2-eB\sigma_z).
\end{eqnarray}
Further, we shall deal only with the two-dimensional case.

Let us first take a look at the supersymmetry formulation of the 
two-dimensional
Pauli equation in the case of the stationary magnetic field
\cite{Grum,Jac,Khare,Gen}. 
Then the solution of the nonstationary equation can be written in the 
form of
$\psi(x,y,t)=e^{-iEt}\psi(x,y)$ 
and for $\psi(x,y)$ we obtain the stationary
Pauli equation
\begin{eqnarray}
&& H\psi(x,y)=E\psi(x,y).
\end{eqnarray}
The two-dimensional Pauli Hamiltonian can be written as follows 
\begin{eqnarray}
&& H=\left(
\begin{array}{cc}
{H_+} & {0}\cr
{0} & {H_-}\cr
\end{array}
\right)
={1 \over 2} \left(
\begin{array}{cc}
{\pi_- \pi_+} & {0} \cr
{0} & {\pi_+ \pi_-} \cr
\end{array}
\right),
\end{eqnarray}
or in supersymmetrical form
\begin{eqnarray}
&& H=\{Q_+, Q_-\}.
\end{eqnarray}
The supercharges $Q_+$, $Q_-$ read
\begin{eqnarray} \label{stand_supercharges}
&& Q_+={\pi_- \sigma_+ \over \sqrt2 },\\
&& Q_-={\pi_+ \sigma_- \over \sqrt2 }, \nonumber 
\end{eqnarray}
where
\begin{eqnarray*}
&& \pi_{\pm}=\pi_x \pm i \pi_y,  \\
&& \sigma_{\pm}={{\sigma_x \pm i \sigma_y} \over 2}.
\end{eqnarray*}

It is convenient to introduce the complex variable $z=x+iy$. 
Then
\begin{eqnarray}
&& \pi_+=-2i{\partial \over \partial z^*} - eA(z,z^*), \\
&& \pi_-=-2i{\partial \over \partial z} - eA^*(z,z^*), \nonumber \\
&& A(z,z^*)=A_x+iA_y. \nonumber
\end{eqnarray}
The introduced supercharges (7) satisfy also the following relations
\begin{eqnarray}
&& (Q_{\pm})^2=0, \\
&& [Q_\pm, H]=0. \nonumber
\end{eqnarray}
These relations together with (6) define the supersymmetry algebra,
which explains the two-fold degeneracy of the 
non-zero energy levels of the electron 
in the two-dimensional inhomogeneous magnetic field $B(x,y)$.

Note that the supercharges $Q_{\pm}$ commute with the Hamiltonian
and in the case of the stationary magnetic field they are the integrals
of motion. 
In the case of the nonstationary magnetic field all relations (5)-(9) are
also true but $Q_{\pm}$ are not the integrals of motion. 
In the next section we shall establish the 
supercharges $\tilde{Q}_{\pm}$ which are the integrals of the 
electron motion in the nonstationary electromagnetic field. 

\section{Supercharges in a nonstationary case}
The supercharges which are the integrals of motion must
satisfy the equation 
\begin{eqnarray}
&& i{\partial \tilde{Q}_{\pm} \over \partial t}+[\tilde{Q}_{\pm},H]=0. 
\end{eqnarray}
This equation was used in \cite{Tkachuk} for the calculation of 
the supercharges
in the case of the nonstationary magnetic field and some results
of paper \cite{Kost} concerning SUSY were reproduced in more simple way.
In this paper we consider the case of the nonstationary axially-symmetric
electromagnetic
field with the vector potential 
\begin{eqnarray}
&& A_x=-{1\over 2}B(t)y +{1\over 2}D(t)x, \\
&& A_y={1\over 2}B(t)x+ {1\over 2}D(t)y. \nonumber 
\end{eqnarray}
Then the magnetic field is
\begin{equation}
B_x=B_y=0, B_z=B(t).
\end{equation}
The components of the electric field consist of two parts
\begin{eqnarray}
&& E_x={1\over 2}{\partial B(t)\over \partial t}y -
{1\over 2}{\partial D(t) \over \partial t}x, \\
&& E_y=-{1\over 2}{\partial B(t)\over \partial t}x -
{1\over 2}{\partial D(t) \over \partial t}y,  \nonumber 
\end{eqnarray}
where the first term is the solenoidal field connected with the
time-varying of the magnetic field, the second term is a potential electric
field of the nonstationary harmonic oscillator. 
Thus, the system considered is
an isotropic two-dimensional nonstationary harmonic oscillator in
the time-varying uniform magnetic field. Note, that the case $D(t)=0$
corresponds to the one considered in \cite{Kost, Tkachuk}.

In complex variables
\begin{equation}
A(z,z^*,t)={1\over 2} a(t)z,
\end{equation}
where
\begin{equation}
a(t)=D(t)+ iB(t)
\end{equation}
and then
\begin{eqnarray}
&& \pi_-=-2i{\partial \over \partial z} - {e a^*(t) \over 2}z^*, \\
&& \pi_+=-2i{\partial \over \partial z^*} - {e a(t) \over 2}z. \nonumber
\end{eqnarray}
The solution of equation (10) $\tilde{Q}_{\pm}$ can be written in the form
similar to (7) 
\begin{eqnarray} \label{supercharges}
&& \tilde{Q}_+={\tilde{\pi}_-\sigma_+ \over \sqrt2}, \\
&& \tilde{Q}_-={\tilde{\pi}_+\sigma_- \over \sqrt2}, \nonumber 
\end{eqnarray}
where
\begin{eqnarray}
&& \tilde{\pi}_-=-f_1(t)2i{\partial \over \partial z} - f_2(t){ea^* \over 
2}z^*, \\
&& \tilde{\pi}_+=-f_1^*(t)2i{\partial \over \partial z^*} - f_2^*(t){ea 
\over 2}z, \nonumber
\end{eqnarray}
$f_1(t)$ and $f_2(t)$ are the unknown functions that will be calculated
further.
For this purpose the following commutation relations are useful
\begin{eqnarray}
&& [\pi_-,\pi_+]=ie(a-a^*)=-2eB, \\
&& [\tilde{\pi}_-,\tilde{\pi}_+]=ie(f_1f_2^*a + f_1^*f_2 a^*), \nonumber \\
&& [\tilde{\pi}_-,\pi_+]=ie(f_1a +f_2a^*), \nonumber \\
&& [\pi_-,\tilde{\pi}_+]=-ie(f_1^*a^* +f_2^*a), \nonumber \\
&& [\pi_-,\tilde{\pi}_-]=[\pi_+,\tilde{\pi}_+]=0. \nonumber
\end{eqnarray}
Substituting $\tilde{Q}_+$ from (17) into (10) we obtain the equation for 
$\tilde{\pi}_-$
\begin{eqnarray}
&& i{\partial \tilde{\pi}_- \over \partial t} + 
\tilde{\pi}_-H_--H_+\tilde{\pi}_-=0.
\end{eqnarray}

Before considering the equation for $f_1$ and $f_2$ note that 
$\tilde \pi_-$
is in a simple way related with the integral of motion. Indeed, 
using the fact that
$$H_+ - H_- = [\pi_-,\pi_+]=-2eB(t),$$
equation (20) can be written in the following form
\begin{equation}
 i{\partial \tilde{\pi}_- \over \partial t} + 
[\tilde{\pi}_-,H_- ]  + 2eB(t)\tilde{\pi}_-=0.
\end{equation}
Let 
\begin{equation}
{\tilde\pi}_-=e^{i2\Omega(t)}I,
\end{equation}
where
$
\Omega(t) =e\int_0^t B(t)dt.
$
Then $I$ satisfies the equation
\begin{equation}
 i{\partial I \over \partial t} + [I,H_- ]  =0.
\end{equation}
Thus, $I$ is the integral of motion of the Hamiltonian $H_-$. It is obvious
that $I$ is also the integral of motion of $H_+$ and thus the
integral of motion of the total Hamiltonian $H$. 
It is worth noting also that
\begin{equation}
{\tilde\pi}_+=e^{-i2\Omega(t)}I^+,
\end{equation}
where
$I^+$ is the integral of motion conjugated to $I$.
The question about the integrals of motion of the nonstationary quantum
mechanical problems was studied in \cite{Mal}.

Now, let us consider the equations for $f_1$ and $f_2$. 
Using the commutation relations (19) and substituting the explicit 
expression for
$\tilde{\pi}_-$ (18) in (20) or (21) we obtain a set of equations for 
$f_1(t)$ and $f_2(t)$
\begin{eqnarray}
&& {\partial f_1 \over \partial t}+ea^*(t)(f_1-f_2)=0, \\
&& {\partial (f_2 a^*(t)) \over \partial t}+ea^*(t)a(t)(f_1-f_2)=0. \nonumber
\end{eqnarray}

The solution of this set of equations can be written as follows
\begin{eqnarray}
&& f_1 = f e^{i\Omega(t)},\\
&& ef_2 a^* = (eD(t)f + {\partial f \over \partial t}) e^{i\Omega(t)}, 
\nonumber
\end{eqnarray}
where $f$ satisfies the second-order differential equation
\begin{equation}
{\partial^2 f \over \partial t^2}=-((eB(t))^2 + e{\partial D(t) 
\over \partial t})f.
\end{equation}

Thus, in our case the problem 
of constructing the supercharges $\tilde{Q}_\pm$ leads to equation  
(27). Solving this equation and using (26)
we obtain $f_1$, $f_2$ and thus supercharges  
(\ref{supercharges}) $\tilde{Q}_\pm$, where
\begin{eqnarray}
&& \tilde{\pi}_-= e^{i \Omega(t)}(-f2i{\partial \over \partial z} - 
({\partial f \over \partial t}+eD(t)f)
{1 \over 2}z^*), \\
&& \tilde{\pi}_+=e^{-i \Omega(t)}(-f^*2i{\partial \over \partial z^*}- 
({\partial f^* \over \partial t}+eD(t)f^*)
{1 \over 2}z). \nonumber
\end{eqnarray} 
In conclusion of this section let us consider
the stationary magnetic and electric fields
$$B(t)=B=const,$$
$$D(t)=Dt, \ \ D=const.$$
Equation (27) in this case can be easily solved
\begin{equation}
 f=c_1e^{-i\omega t} + c_2 e^{i\omega t}, 
\end{equation}
where $\omega=\sqrt{(eB)^2+eD}$, $c_1, c_2$ are arbitrary constants.

We have two linearly independent solutions.  
Choosing $c_1=1$, $c_2=0$ gives the following supercharges
\begin{eqnarray}
&& {\tilde Q}_+={1 \over \sqrt 2}e^{i(\omega_0 -\omega)t}
(-2i{\partial \over \partial z} -
(eDt - i\omega){1\over 2}z^*)\sigma_+,\\ \nonumber
&& {\tilde Q}_-={1 \over \sqrt 2}e^{-i(\omega_0 -\omega)t}
(-2i{\partial \over \partial z^*} -
(eDt + i\omega){1\over 2}z)\sigma_-,
\end{eqnarray} 
where $\omega_0=eB$.
The second linearly independent solution $c_1=0$, $c_2=1$ 
gives 
\begin{eqnarray}
&& {\tilde Q}_+={1 \over \sqrt 2}e^{i(\omega_0 +\omega)t}
(-2i{\partial \over \partial z} -
(eDt + i\omega){1\over 2}z^*)\sigma_+,\\ \nonumber
&& {\tilde Q}_-={1 \over \sqrt 2}e^{-i(\omega_0 +\omega)t}
(-2i{\partial \over \partial z^*} -
(eDt - i\omega){1\over 2}z)\sigma_-.
\end{eqnarray} 

Note that even in the constant electromagnetic field supercharges
(30) and (31) depend on the time $t$. When the electric field is
equal to zero $D=0$ the supercharges (30) result in the well known time 
independent supercharges (7). Supercharges (31) in the case of
$D=0$ give new supercharges which are connected with (7)
in the following way 
\begin{eqnarray}
&& \tilde{Q}_+=e^{i2 \omega_0 t}Q_+(-B), \\
&& \tilde{Q}_-=e^{-i2 \omega_0 t} Q_-(-B), \nonumber
\end{eqnarray}
where $Q_\pm(-B)$ are supercharges (7) with the opposite directed 
magnetic field. 

\section{The algebra of supersymmetry}
The supercharges $\tilde{Q}_\pm$ are the integrals of motion of $H$ 
and they fulfil the superalgebra
\begin{eqnarray}
&& 
\{\tilde{Q}_+, \tilde{Q}_-\} = \tilde{H}, \qquad (\tilde{Q}_\pm)^2=0; 
\end{eqnarray}
where
$$ \tilde{H}= {1\over2}\pmatrix{\tilde{\pi}_-\tilde{\pi}_+ 
& 0 \cr 0 & \tilde{\pi}_+\tilde{\pi}_- \cr }$$ 
is also the integral of motion.  $\tilde H$ can be treated
as a new Hamiltonian and is a time-dependent extension of the
usual Hamiltonian $H$.

One more integral of motion results from the axial symmetry. It is 
a $z$-component of the angular momentum
\begin{equation}
L_z = z {\partial \over \partial z} - z^* {\partial \over \partial z^*},
\end{equation}
which satisfies the commutation relation
\begin{equation}
[{\tilde \pi}_\pm, L_z]=\mp{\tilde \pi}_\pm.
\end{equation}
It is obvious that $S_z={\sigma_z}/2$ is also an integral of motion.
The algebra of SUSY can be extended by $L_z$ and $S_z$ which satisfy
the following relations
\begin{eqnarray}
&& [{\tilde Q}_\pm,L_z]=\pm{\tilde Q}_\pm,
\ \ \ \  [{\tilde Q}_\pm,S_z]=\mp{\tilde Q}_\pm,\\ \nonumber
&& [L_z,{\tilde H}]=[S_z,{\tilde H}]=[S_z,L_z]=0.
\end{eqnarray}
The total moment $J=L_z + S_z$ commutes with all the generators of 
the algebra.

Note, that using (19) and (26) we obtain
\begin{equation}
 [ \tilde{\pi}_- , \tilde{\pi}_+] = i(f {\partial f^* \over \partial t}-
 f^* {\partial f \over \partial t}),
\end{equation}
where $f$ and $f^*$ are two linear independent solutions of
equation (27). The Wronskian of equation (27) is constant.
We choose to scale the solutions so that the Wronskian is
$$f {\partial f^* \over \partial t}-
 f^* {\partial f \over \partial t}=-2i, $$
then
$$
[ \tilde{\pi}_- , \tilde{\pi}_+] =2. 
$$

Let us introduce the operators of creation and annihilation
$${\tilde b}_\pm={{\tilde\pi}_\pm \over \sqrt 2}$$
satisfying the commutation relation $[{\tilde b}_-,{\tilde b}_+]=1$.
It is convenient to introduce the operators of creation and annihilation
$b_\pm$ which are the integrals of motion.
Using (22) and (24) we have
\begin{equation}
b_\pm =e^{\pm i2\Omega(t)} {\tilde b}_\pm.
\end{equation}
It is obvious that 
\begin{equation}
[b_- , b_+] =1, 
\end{equation}
and using (35)
\begin{equation}
[b_\pm ,L_z ]=\mp b_\pm.
\end{equation}
The new Hamiltonian $\tilde H$ can be written as follows
\begin{equation}
{\tilde H}={\tilde b}_+{\tilde b}_- +\sigma_+ \sigma_- ={\tilde b}_+
{\tilde b}_- + s_z + 1/2=b_+b_- + s_z + 1/2. 
\end{equation}
In the new notation the supercharges are
\begin{equation}
Q_+={\tilde b}_-\sigma_+=e^{i2 \Omega(t)} b_-\sigma_+, 
\ \ Q_-={\tilde b}_+\sigma_- =e^{-i2 \Omega(t)} b_+\sigma_-.
\end{equation}

\section{Eigensolutions of $\tilde H$ and representation of the algebra}
Let us choose the orthogonal basis $|n,m,s>$, where $n=0,1,2,...$
are the eigenvalues of $b_+b_-$,
$m$ is the eigenvalue of $L_z$,
$s=\pm 1/2$ is the eigenvalue of $S_z$.

The action 
of the operators on the basis is the following
\begin{eqnarray}
&& {\tilde H}|n,m,s>=(n+s+1/2)|n,m,s>,\\ \nonumber
&& L_z |n,m,s>=m|n,m,s>,\\ \nonumber
&& S_z |n,m,s>=s|n,m,s>,\\ \nonumber
&&b_-|n,m,s>=\sqrt{n}|n-1,m-1,s>, \\ \nonumber
&&b_+|n,m,s>=\sqrt{n+1}|n+1,m+1,s>, \\ \nonumber
&& {\tilde Q}_+ |n,m,s>= \delta_{-1/2,s}e^{i2\Omega (t)} \sqrt{n}
|n-1,m-1,1/2>, 
\\ \nonumber
&& {\tilde Q}_- |n,m,s>= \delta_{1/2,s}e^{-i2\Omega (t)} \sqrt{n+1}
|n+1,m+1,-1/2>. 
\end{eqnarray}
The considered basis can be obtained by applying the operator of 
creation $b_+$ to the ground state
\begin{equation}
|n,m,s>={1\over\sqrt{n!}} (b_+)^n |0,m-n,s>.
\end{equation}
The ground state is defined by the equation
\begin{equation}
b_-|0,m,s>=0.
\end{equation}
The solution of this equation taking into account spin variables
is the following
\begin{equation}
|0,m,s>=C_s(t) \chi_s (z^*)^{-m} \exp{i\over 4}(eD(t) +
{1\over f(t)} {\partial f(t) \over \partial t})|z|^2,
\end{equation}
where 
$$
\chi_{1/2}=\left(
\begin{array}{c}
1\\
0
\end{array}
\right), \ \ \
\chi_{-1/2}=\left(
\begin{array}{c}
0\\
1
\end{array}
\right)
$$
are the eigenstates of the spin operator $S_z$, from the condition 
that the wave function does not have a pole at zero $m \le 0$,
$C_s(t)$ is a certain function
of time. We may choose this function in the way that eigenstates of
$\tilde H$ satisfy the nonstationary Pauli equation. It is easy
to check that the wave function (46) with
\begin{equation}
C_s(t)=C (f(t))^{m-1}e^{i(m+2s)\Omega(t)}
\end{equation}
indeed satisfies the nonstationary Pauli equation. Here $C$ is 
a constant of normalization of wave function (46).

It is known
that applying an integral of motion to the solution of wave equation
yields a function that itself is the solution of the same wave
equation (see for example \cite{Kost, Mal}). Thus, because $b_+$
is the integral of motion, the wave function $|n,m,s>$ calculated
by (44) is also a solution of the nonstationary Pauli equation.
For more details of constructing the exact solution of the
nonstationary Schr\"odinger equation with the help of the 
dynamical symmetry see \cite{Mal}.
Note also that $\tilde H$ has equidistant two-fold degenerated 
eigenvalues connected with SUSY, except for the unique ground state.

\section{Conclusion}

Originally, the SUSY of the Pauli equation was established
in the case of the stationary magnetic field without the electric one. 
The supercharges in this case do not depend on time.
In \cite{Kost} it was shown that SUSY can be introduced also
in the case of the nonstationary magnetic field.

In the present paper 
the SUSY of the Pauly equation is introduced in the case of both 
the nonstationary
magnetic and electric fields. The supercharges are obtained as
integrals of motion of the Hamiltonian.
It is worth noting that although the supercharges ${\tilde Q}_\pm$ and
the new Hamiltonian $\tilde H$ are the integrals of motion, there is
implicit time dependence in them.
In the case of the stationary magnetic and electric fields the
obtained supercharges remain implicitly time dependent (see (31)). 
Only in the case of the stationary magnetic field without the electric one
there exist the supercharges which do not contain time in implicit form.
Note also that the obtained algebra can be used for generating solutions of
the nonstationary Pauli equation as is shown in section 5.
\pagebreak

\end{document}